\documentclass[twocolumn,prc,nofootinbib,aps,final,amsmath,amssymb,10pt,superscriptaddress,floatfix,letterpaper,showpacs]{revtex4-1}
\usepackage{eurosym}
\usepackage{amsmath}
\usepackage{amsfonts}
\usepackage{amssymb}
\usepackage[bookmarksnumbered,colorlinks,bookmarks,breaklinks=true,linkbordercolor={1 1 1},linktocpage,citecolor=blue]{hyperref}
\usepackage{graphicx,color}
\usepackage{subfigure}
\usepackage{dcolumn}
\usepackage{bm}

\usepackage{ulem}

\def\etal{\textit{et al.~}}
\def\etals{\textit{et al.}}
\def\ql{``}
\def\qr{''\hspace{0.5mm}}
\def\qrs{''}
\def\PB{$^{208}$Pb}

\begin{document}

\title{Dispersive optical model description of nucleon scattering on Pb--Bi isotopes}

\author{Xiuniao Zhao}
\affiliation{Graduate School of China Academy of Engineering Physics, Beijing 100088, China}

\author{Weili Sun}
\affiliation{Institute of Applied Physics and Computational Mathematics, Beijing 100094, China}
\email{sun$\_$weili@iapcm.ac.cn}

\author{R.~Capote}
\affiliation{NAPC--Nuclear Data Section, International Atomic Energy Agency, Vienna A-1400, Austria}
\email{r.capotenoy@iaea.org}

\author{E.Sh.~Soukhovitski\~{\i}}
\affiliation{Joint Institute for Energy and Nuclear Research, Minsk-Sosny 220109, Belarus}

\author{D.S.~Martyanov}
\affiliation{Joint Institute for Energy and Nuclear Research, Minsk-Sosny 220109, Belarus}

\author{J.M.~Quesada}
\affiliation{Departmento de F\'isica At\'omica, Molecular y Nuclear, Universidad de Sevilla, Apartado 1065, Sevilla E-41080, Spain}

\vspace{10pt}

\begin{abstract}
A recently derived dispersive optical model potential (DOMP) for $^{208}$Pb is extended to consider the non-locality in the real potential and the shell-gap in the definition of the nuclear imaginary potentials near the Fermi energy. The modified DOMP improves the simultaneous description of nucleon scattering on $^{208}$Pb and of the $^{208}$Pb particle-hole bound states. The new potential is shown to give a very good description of nucleon scattering data on near-magic targets $^{206,207}$Pb  and $^{209}$Bi.
\end{abstract}

\maketitle


\section{Introduction}

The nuclear optical model has been comprehensively applied to analyse the elastic scattering of pions, nucleons and heavier particles by nuclei over a wide range of energies \cite{Hodg63,Hodg71,Hodg97}. Requirement of causality, namely that the scattering wave is not emitted before the incident wave arrives \cite{mangsa86}, led to the need to consider dispersion effects in the nuclear scattering, and allowed to combine the optical model potential and the shell model potential into a dispersive optical model potential (DOMP)~\cite{masa86}. The DOMP combined both nuclear reaction ($E>0$) and nuclear structure ($E<0$) information to minimize the number of parameters and improve the predictive capabilities of relevant observables.


Pioneering work on DOM potentials for strongly deformed nuclei was the contribution of Romain and Delaroche \cite{rode97} devoted to
the analysis of the nucleon scattering data on $^{181}$Ta and tungsten isotopes. An explicit treatment of the non-locality of the surface imaginary
potential and of the \ql Hartree-Fock\qr (HF) potential was introduced following Perey-Buck recipes \cite{PereyBuck62}.

Mahaux and Sartor suggested in 1991 \cite{masa91rev,masa91} that the absorptive potential will be asymmetric at large positive and negative energies with respect to the Fermi energy
$E_{F}$. The DOM analysis of neutron scattering on $^{27}$Al \cite{mocaqu:02} showed the importance of the asymmetry of the volume absorptive potential
and the corresponding dispersive contributions to describe ${\sigma}_{T}$ data for energies above 100 MeV.

Many studies have also dealt with nucleon scattering on near magic nuclei. A global spherical potential for nucleon induced reactions derived by Koning and Delaroche
\cite{Koning03} used local dispersive OMPs as starting point \cite{ko00}. Recently a global dispersive spherical potential for neutron induced reactions
was derived by Morillon and Romain \cite{moro:04}, where an explicit nonlocal HF-like potential was used; bound-state data were also studied~\cite{moro:06}.

Dispersive optical model has been extensively developed by Washington University (St. Louis) researchers to study nucleon scattering on magic and near magic nuclei as reviewed recently
by Dickhoff and Charity \cite{dich:2018}. The need to introduce asymmetric imaginary volume potentials far from the Fermi energy was confirmed in Ref.~\cite{Charity:06} and led to an improved  description of spectroscopic factors of the bound states \cite{Mueller:11}. An energy gap called $E_P$ near the Fermi energy was introduced in Ref.~\cite{Mueller:11} to describe elastic nucleon scattering data on magic nuclei. Additionally, the importance of the spatial non-locality in the DOM potential, including both the real and imaginary parts, was highlighted in Refs.~\cite{Wald:11,Dick:10,Mueller:11,Mahz:14,Duss:14,Mahz:17} to describe both the nucleon scattering as well as bound-state data. Non-locality in the DOM was also shown to have a large impact on calculated (p,d) transfer cross sections \cite{Ross:15}.

Phenomenological local DOM potentials following the Lane formulation \cite{la62a,la62b} have been developed by authors \cite{Souk:05,casoqu05,Capote:08,ripl3,Li:2013,Souk:16,Sun17} and mostly applied to describe nucleon scattering on well deformed target nuclei using a coupled-channel formalism. Calculated scattering cross sections included quasi-elastic (p,n) scattering data, \textit{e.g.}, see Ref.~\cite{Quesada:2007}. Those potentials very accurately describe available experimental data of nucleon scattering from keV up to 150--200~MeV of incident nucleon energy. However, deformed nuclei do not have bound-state experimental data available as the bound states are very fragmented due to the deformation.

The analysis of nucleon scattering of $^{208}$Pb by DOMP  was recently undertaken \cite{Sun17}. The DOMP from Ref.~\cite{Sun17} was also used to test the derived DOMP at negative energies using our methodology \cite{Zhao:2019}. Calculated DOMP energies of the particle-hole bound states were compared to other calculated values \cite{moro:04,moro:06} as well as to the existing experimental data~\cite{johoma87}. Some inconsistencies in the data description were found in Ref.~\cite{Zhao:2019} including problems to describe accurately the total cross sections in the region from 5 up to 10 MeV and, at the same time, achieve a nice description of the bound-state data.

In this work, some of the physical ideas advanced by Mahaux and Sartor \cite{masa91rev}, the CEA Bruy\`{e}res-le-Ch\^{a}tel group \cite{rode97,moro:04,moro:06}, and the Washington University (St Louis) group \cite{Charity:06,Mueller:11} will be tested using our phenomenological DOMP framework to study the impact on calculated observables. Our main goal is to derive a new Lane consistent potential for lead and bismuth isotopes that reproduce very well both scattering and bound-state data.

\begin{table*}[!ht]
\vspace{-3mm}
\caption{Dispersive optical-model potential parameters for nucleon induced reactions on lead and bismuth isotopes.}
\vskip 2mm \renewcommand{\arraystretch}{1.15} \tabcolsep 5pt
\centerline{\begin{tabular*}{0.91\textwidth}{ccccc}
  \hline
  \hline
 & Volume & Surface & Spin-orbit & Coulomb \\
\hline
     & $V_{0}$=81.5+0.0292(A-208)~MeV & & $V_{\rm{so}}$=7.61~MeV & $C_{\rm{Coul}}$=1.288~MeV \\
Real & $\beta$=0.912~fm & dispersive & $\lambda_{\rm{so}}$=0.006~MeV$^{-1}$ &  \\
Potential & $C_{\rm{viso}}$=29.35~MeV & ($\Delta V_{s}$) & + dispersive ($\Delta V_{so}$)  \\
 & + dispersive ($\Delta V_{v}$) & &  \\
\hline
 & $A_{\rm{v}}$=12.81~MeV & $W_{0}$=19.66~MeV & $W_{\rm{so}}$=-3.1~MeV &  \\
Imaginary & $B_{\rm{v}}$=65.56~MeV& $B_{\rm{s}}$=8.99~MeV & $B_{\rm{so}}$=160~MeV &  \\
Potential & $E_{\rm{a}}$=56~MeV & $C_{\rm{s}}$=0.025~MeV$^{-1}$ & &  \\
 & $\alpha=0.12$~MeV$^{1/2}$ & $C_{\rm{wiso}}$=50.71~MeV & &  \\
\hline
Potential & $r_{\rm{HF}}$=1.226-0.00176(A-208) & $r_{\rm{s}}$=1.1858+0.03418(A-208) & $r_{\rm{so}}$=1.194 & $r_{\rm{c}}$=1.27 \\
Geometry & $a_{\rm{HF}}$=0.647+0.002417(A-208) & $a_{\rm{s}}$=0.6195 & $a_{\rm{so}}$=0.6426 & $a_{\rm{c}}$=0.671 \\
(fm)& $r_{\rm{v}}$=1.321 & & &  \\
& $a_{\rm{v}}$=0.6267-0.00658(A-208) & & &  \\
\hline
\end{tabular*}}
\label{tabI}
\vspace{-4mm}
\end{table*}

\begin{table}[!ht]
\vspace{-3mm}
\caption{The average particle (hole) single-particle energies $E_{\rm{P}}$ (for neutrons and protons) in MeV for nucleon induced reaction on selected targets.}
\vskip 0mm \renewcommand{\arraystretch}{1.25} \tabcolsep 12pt
\centerline{\begin{tabular*}{0.5\textwidth}{ccccc}
  \hline
  \hline
 & $^{206}$Pb & $^{207}$Pb & $^{208}$Pb & $^{209}$Bi\\
 \hline
$E_{\rm{P}}$(n)& -6.75 & -6.74 & -3.95 & -4.62\\
$E_{\rm{P}}$(p)& -3.57 & -3.72 & -3.81 & -3.81\\
\hline
\end{tabular*}}
\label{tabII}
\vspace{-4mm}
\end{table}

\section{DISPERSIVE SPHERICAL OPTICAL MODEL POTENTIAL}
A dispersive optical model is defined by energy-dependent real $V_i$ ($i=HF,v,s,C,so$) and imaginary $W_i$ ($i=v,s,so$) functionals for the so-called \ql Hartree-Fock\qr (HF), volume (v), surface (s), Coulomb (C) and spin-orbit (so) potentials, respectively and also by the corresponding dispersive contributions to the real potential $\Delta V_v$, $\Delta V_s$, and $\Delta V_{so}$ which are calculated analytically from the corresponding imaginary potentials \cite{CPC,PRC-disp,Souk:16}. The general formulation of the Lane-consistent spherical dispersive optical potential has been published previously (\textit{e.g.}, see Eqs.(1)--(3) in Ref.~\cite{Zhao:2019}), and is not repeated here. Note that our formulation considers the Coulomb corrections in all orders through an effective energy shift in the potential definition, \textit{i.e.}, the effect of Coulomb interaction on the nuclear interaction is not an averaged energy-independent constant as usually done (\textit{e.g.}, see Koning-Delaroche potential definition \cite{Koning03}).

It is well known (see \textit{e.g.}, Ref.~\cite{dich:2018}) that the real mean-field potential $V_{\rm{HF}}(\bf{r},\bf{r}')$ is non-local and energy independent. A parametrization of such nonlocal potential was postulated by Perey and Buck to be of Gaussian type~\cite{PereyBuck62}:
\begin{equation}
V_{\rm{HF}}(\bf{r},\bf{r}')=V(\bf{r})\exp{(-|\bf{r}-\bf{r}'|^2/\beta^2)}, \label{PB-nonlocal}
\end{equation}
where the parameter $\beta$ is a non-locality range given in fermi. The local energy approximation of such non-local potential~\cite{PereyBuck62} then results in the following implicit equation:
\begin{equation}
V_{\rm{HF}}(E)=A_{HF}\exp{(-\frac{\mu \beta^2}{(\hbar c)^2}[E+V_{\rm{HF}}(E)])}. \label{PB-approx}
\end{equation}
Note that both $A_{HF}$ and the potential $V_{\rm{HF}}(E)$ in Eq.~\eqref{PB-approx} are assumed to be positive. To obtain the potential depth $V_{\rm{HF}}$(E) at a given energy $E$ it is necessary to solve the Eq.~\eqref{PB-approx} by iterations\footnote{Solution of Eq.~\eqref{PB-approx} can be expressed explicitly through the special function Lambert $W$ (\textit{a.k.a.}, product logarithm) as: \newline
$V_{\rm{HF}}(E)= \frac{W[A_{HF}\lambda\exp{(-\lambda E)}]}{\lambda}$,
where $\lambda\equiv \frac{\mu\beta^2}{(\hbar c)^2}$.}.
Note that both $A_{HF}$ and $\beta$ are independent of iterations on $V_{\rm{HF}}$ for a given energy $E$. The reduced mass $\mu$ in the formula is calculated using relativistic kinematics and, therefore, is also a function of the incident nucleon energy $E$. The isospin dependence of the potential (the Lane term \cite{la62a,la62b}) was considered in real $V_{HF}(E)$ and imaginary surface $W_{s}(E)$ potentials as follow,
\begin{align}
A_{HF}& =V_{0}\left[ 1+(-1)^{Z^{\prime }+1}\frac{C_{viso}}{V_{0}}\frac{N-Z}{A%
}\right]  \label{AHF} \\
A_{s}& =W_{0}\left[ 1+(-1)^{Z^{\prime }+1}\frac{C_{wiso}}{W_{0}}\frac{N-Z}{A}%
\right]
\end{align}%
where $V_{0}$, $C_{viso}$, $W_{0}$ and $C_{wiso}$ are undetermined constants. Many authors found that the imaginary volume potential does not depend on the
isospin. The isospin constants $C_{viso}$ and $C_{wiso}$ should be determined mainly using quasi-elastic (p,n) scattering data.

The energy dependencies for the imaginary volume term $W_{\rm{v}}$, the imaginary surface term $W_{\rm{s}}$ and the spin-orbit imaginary term $W_{\rm{so}}$ are taken as the ones suggested by Brown and Rho~\cite{Brown81}, Delaroche \etal\cite{dewara89}, and Koning \etal\cite{Koning03}, respectively. The imaginary potentials used in all our studies so far are local ones. Some groups advocate the need to consider non-local imaginary potentials~\cite{rode97,Dick:10,Mahz:14}, but this is deferred to future works.
In this work, following Mahaux \etal\cite{masa91rev} and Molina \etal\cite{mocaqu:02}, a modified definition for the imaginary part of the OMP is taken as follows:
\begin{equation}
W_{\rm{v}}(E)=\left\{\begin{array}{cl}
0&E_{\rm{F}}<E<E_{\rm{P}}\\
A_{\rm{v}}\frac{(E-E_{\rm{P}})^{2}}{(E-E_{\rm{P}})^{2}+(B_{\rm{v}})^{2}}&E>E_{\rm{P}}
\end{array} \right.
\end{equation}
\begin{equation}
W_{\rm{s}}(E)=\left\{\begin{array}{cl}
0&E_{\rm{F}}<E<E_{\rm{P}}\\
A_{\rm{s}}\frac{(E-E_{\rm{P}})^{2}}{(E-E_{\rm{P}})^{2}+(B_{\rm{s}})^{2}} \times \\
   \;\;\;\;\;\; \exp{(-C_{\rm{s}}|E-E_{\rm{P}}|)}&E>E_{\rm{P}}
\end{array} \right.
\end{equation}
\begin{equation}
W_{\rm{so}}(E)=\left\{\begin{array}{cl}
0&E_{\rm{F}}<E<E_{\rm{P}}\\
W_{\rm{so}}\frac{(E-E_{\rm{P}})^{2}}{(E-E_{\rm{P}})^{2}+(B_{\rm{so}})^{2}}&E>E_{\rm{P}}
\end{array} \right..
\end{equation}

The imaginary part of the DOM potential is assumed to be zero inside the shell gap $\Delta$, which is related to the average energy of the single-particle (-hole) states $E_{\rm{P}}$ as $\Delta=2(E_{\rm{P}}-E_{\rm{F}})$. Obviously, there are no states in the shell-gap, therefore we have to set the absorption to zero. A similar definition of the shell gap was employed in Refs.~\cite{Charity:06,Mueller:11}. Both $E_{\rm{P}}$ and $E_{\rm{F}}$ are different for neutron and proton induced reactions. For nuclei far from magic $E_{\rm{P}}$ is approximately equal $E_{\rm{F}}$, therefore the shell-gap is zero and can be neglected. The symmetry condition $W(2E_{\rm{F}}-E)=W(E)$ is used to extend the imaginary part of the OMPs for energies below the Fermi energy. This analytical extension is needed for the calculation of the dispersive corrections.
\begin{figure*}[!tbh]
\centering
\subfigure[~Real and imaginary potential depths for volume, surface, and spin-orbit potentials and dispersive contributions.]
{\includegraphics[width=0.48\textwidth]{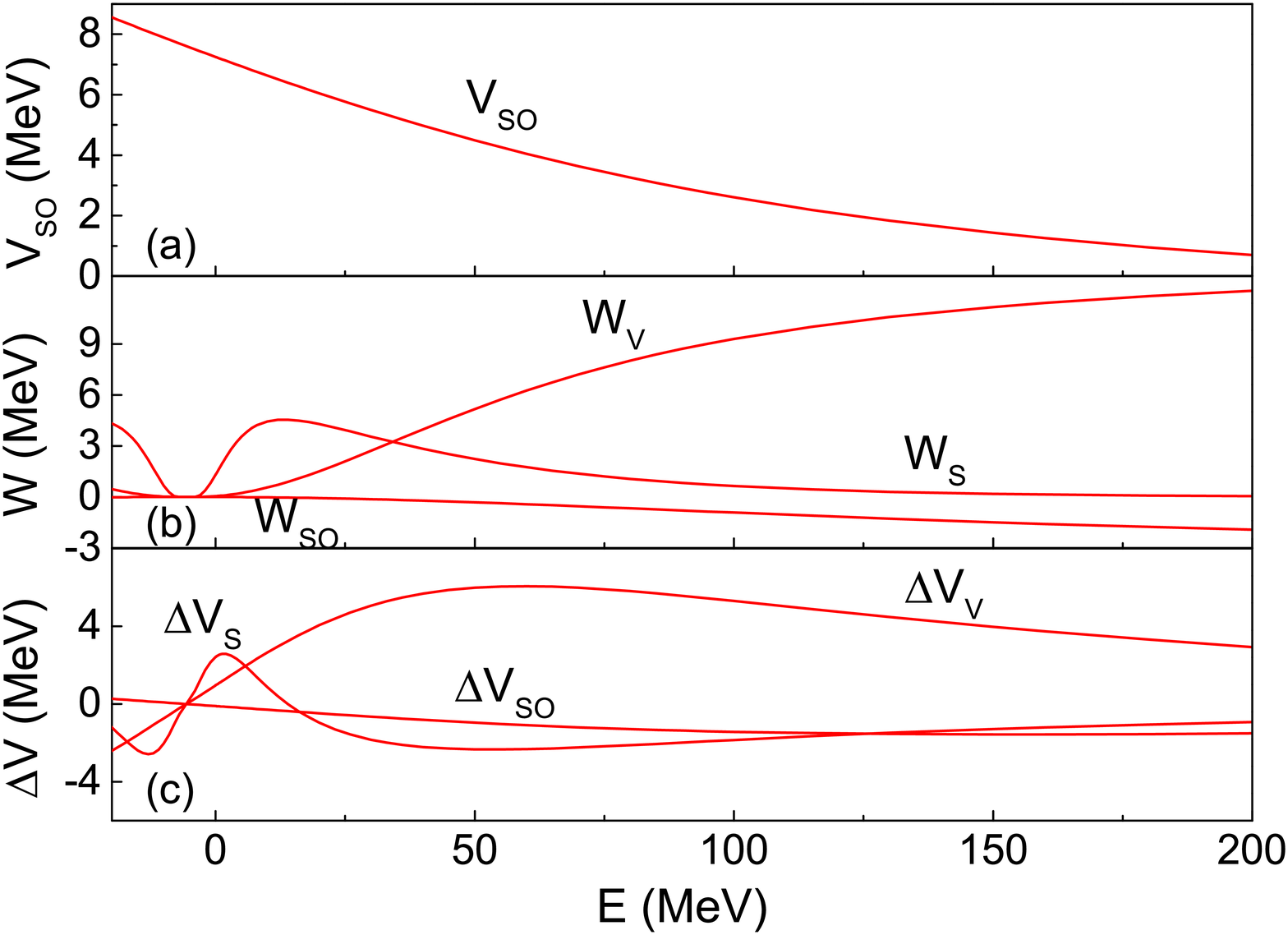}}
\subfigure[~\ql Hartree-Fock\qr potential depth as a function of energy from Ref.~\cite{Zhao:2019} (previous) compared to the present one given by Eq.~\eqref{PB-approx}.]
{\includegraphics[width=0.48\textwidth]{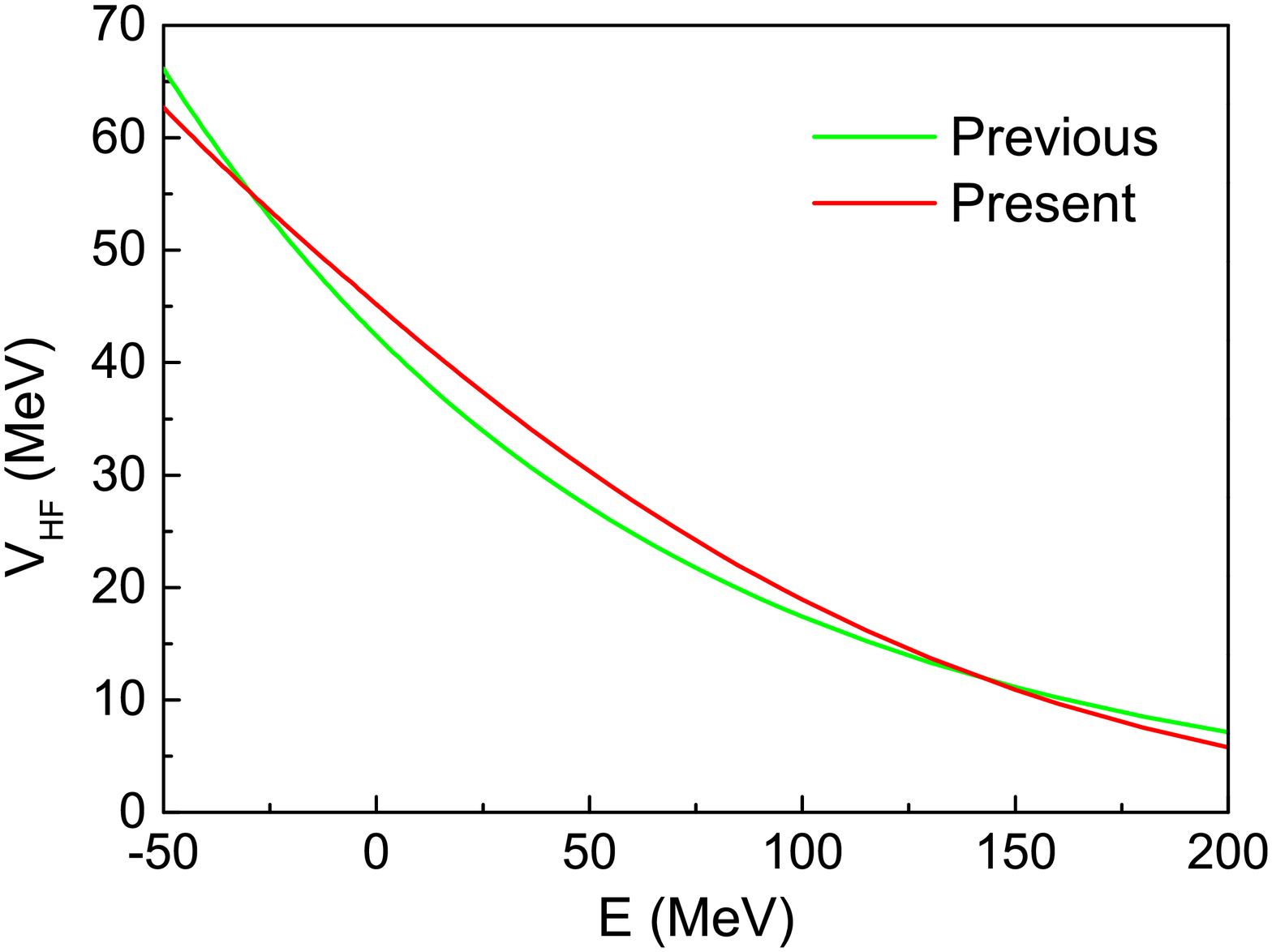}}
\vspace{-2mm}
\caption{DOMP depths and dispersive contributions as a function of $E$ for the n+$^{208}$Pb reaction between -50 and 200~MeV.}
\label{fig1}
\vspace{-4mm}
\end{figure*}

Asymmetric absorptive potentials were used in many analysis of DOMPs derived on different targets \cite{Souk:05,casoqu05,Capote:08,ripl3,Li:2013,Souk:16,Sun17}. Following Mahaux and Sartor \cite{masa91}, the assumption that the imaginary potential $W_{v}(E)$ is symmetric about $E=E_{F}$ (according to equation $W(2E_{F}-E)=W(E)$) is modified above some fixed energy $E_a$, which is expected to be close to 60 MeV, but it is treated as a parameter.

\begin{figure}[!tbh]
\centering
\includegraphics[width=0.95\columnwidth]{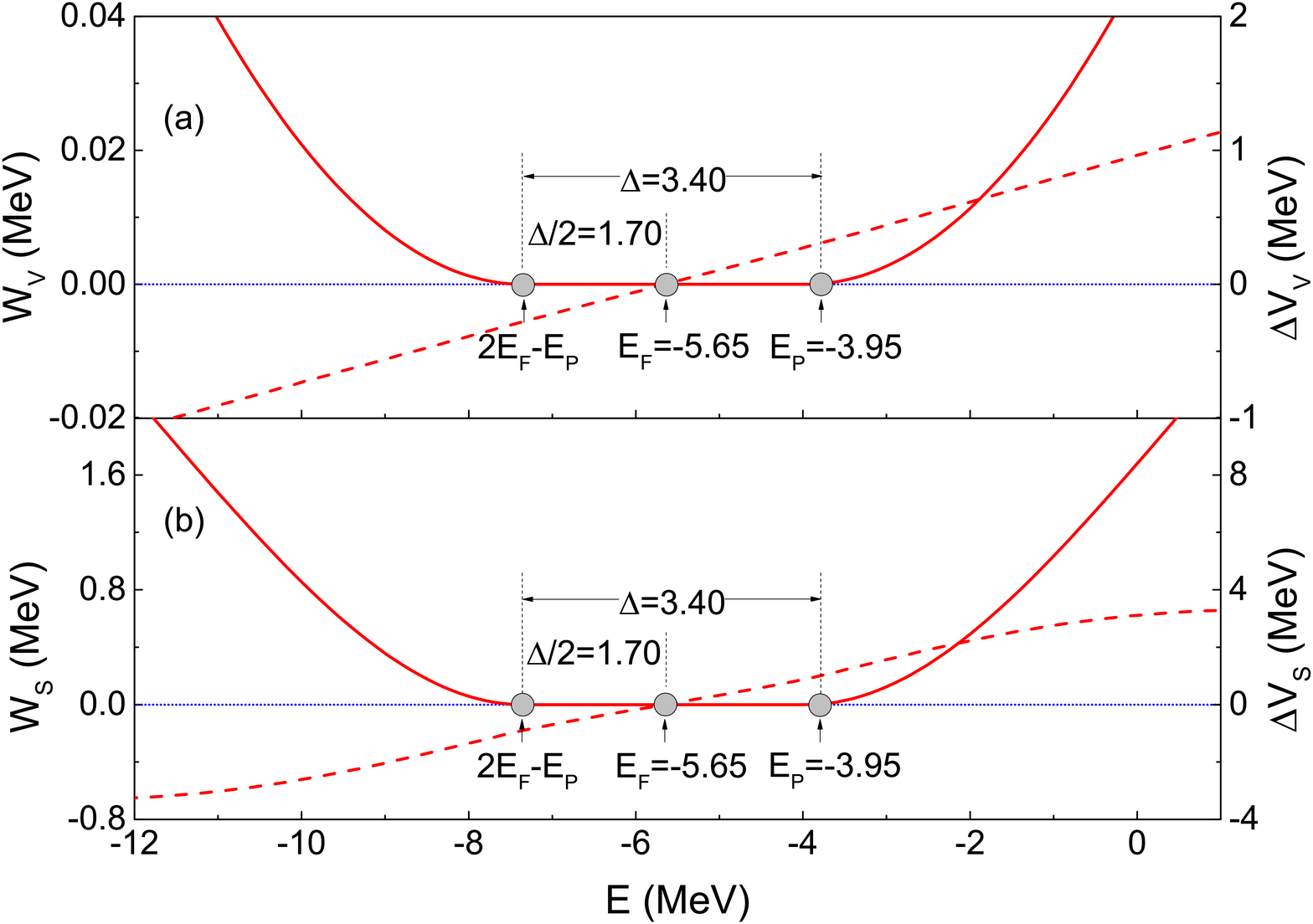}
\caption{Energy dependence of the depths of volume, surface imaginary potential (solid curve) and dispersion correction terms (dashed curve) near the Fermi energy $E_{\rm{F}}$ calculated for the n+$^{208}$Pb reaction. The effect of the assumed shell-gap $\Delta=2(E_{\rm{P}}-E_{\rm{F}})=3.4$~MeV on the imaginary potentials is clearly seen.}
\label{fig2}
\vspace{-4mm}
\end{figure}
\begin{figure}[!thb]
\centering
\includegraphics[width=0.89\columnwidth]{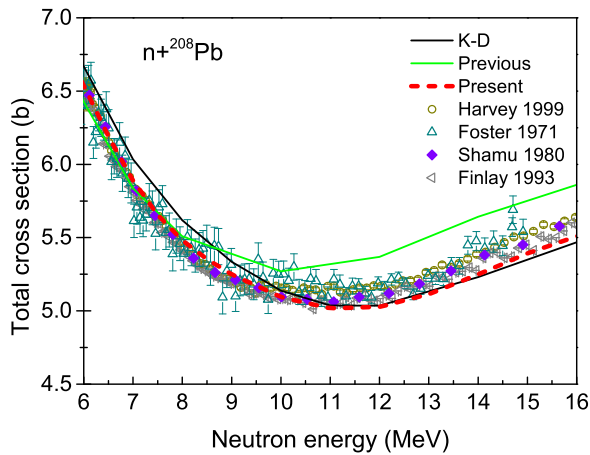}
\vspace{-2mm}
\caption{Comparison of the calculated total cross section for the n+$^{208}$Pb reaction with measurements. Calculations using the Koning-Delaroche \cite{Koning03}, the DOMP from our previous work \cite{Zhao:2019}, and the current DOMP are shown. Experimental data are taken from EXFOR \cite{EXFOR}, Refs.~\cite{Harvey99,FosterJR,Shamu80,Finlay93}.}
\label{fig3}
\end{figure}
\begin{figure*}[!thb]
\vspace{-3mm}
\centering
\includegraphics[width=0.84\textwidth]{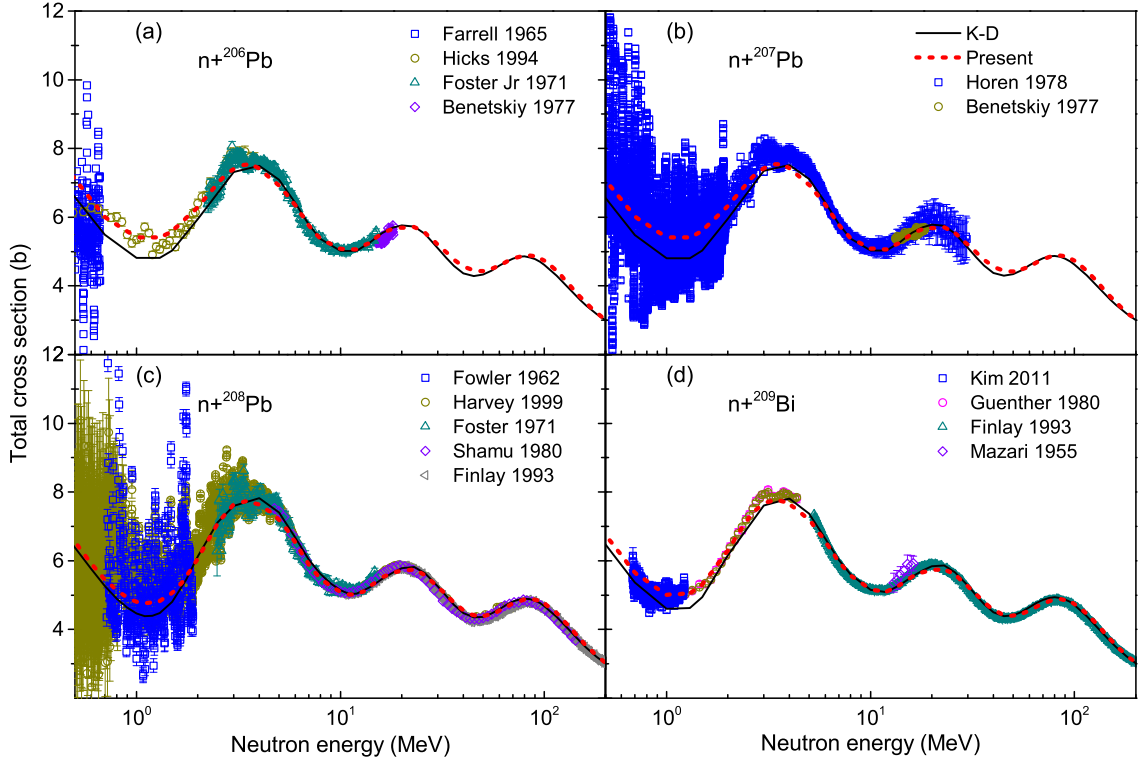}
\caption{Comparison of total cross section for n+$^{206}$Pb, $^{207}$Pb, $^{208}$Pb and $^{209}$Bi reactions with measurements, as well as the results of Koning-Delaroche calculations\cite{Koning03}. Experimental data are taken from Refs.~\cite{Farrell65,Hicks94,Benetskiy77,Horen78,Fowler62,Harvey99,FosterJR,Shamu80,Finlay93,Kim11,Guenther80,Mazari55}.}
\label{fig4}
\vspace{-2mm}
\end{figure*}

\begin{figure*}[t]
\vspace{-3mm}
\centering
\includegraphics[width=0.90\textwidth]{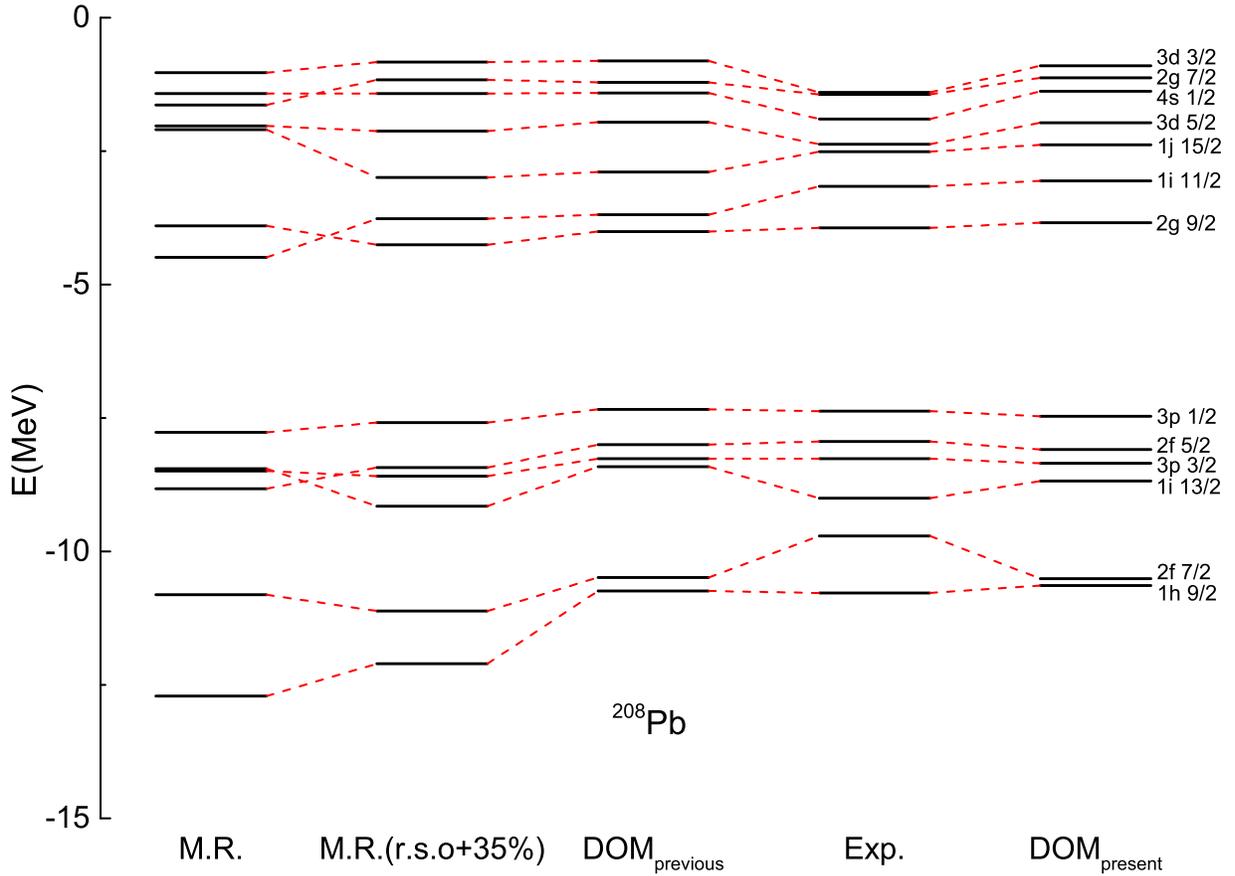}
\caption{Neutron single-particle (hole) energies in $^{208}$Pb, the first and second columns display the results from Ref.~\cite{moro:06}, the third column -- Ref.~\cite{Zhao:2019}, the fifth column -- current work. In the fourth column the experimental values taken from Ref.~\cite{johoma87} are shown. Note that the $E_F\approx -5.6$~MeV and it defines the $N=126$ shell. Levels below $-5$~MeV are hole levels, above are particle levels.}
\label{fig5}
\vspace{-2mm}
\end{figure*}
\begin{figure*}[!tbh]
\vspace{-2mm}
\centering
\subfigure[~$^{206}$Pb and $^{207}$Pb (n,n) angular distributions.]
{\includegraphics[width=0.48\textwidth]{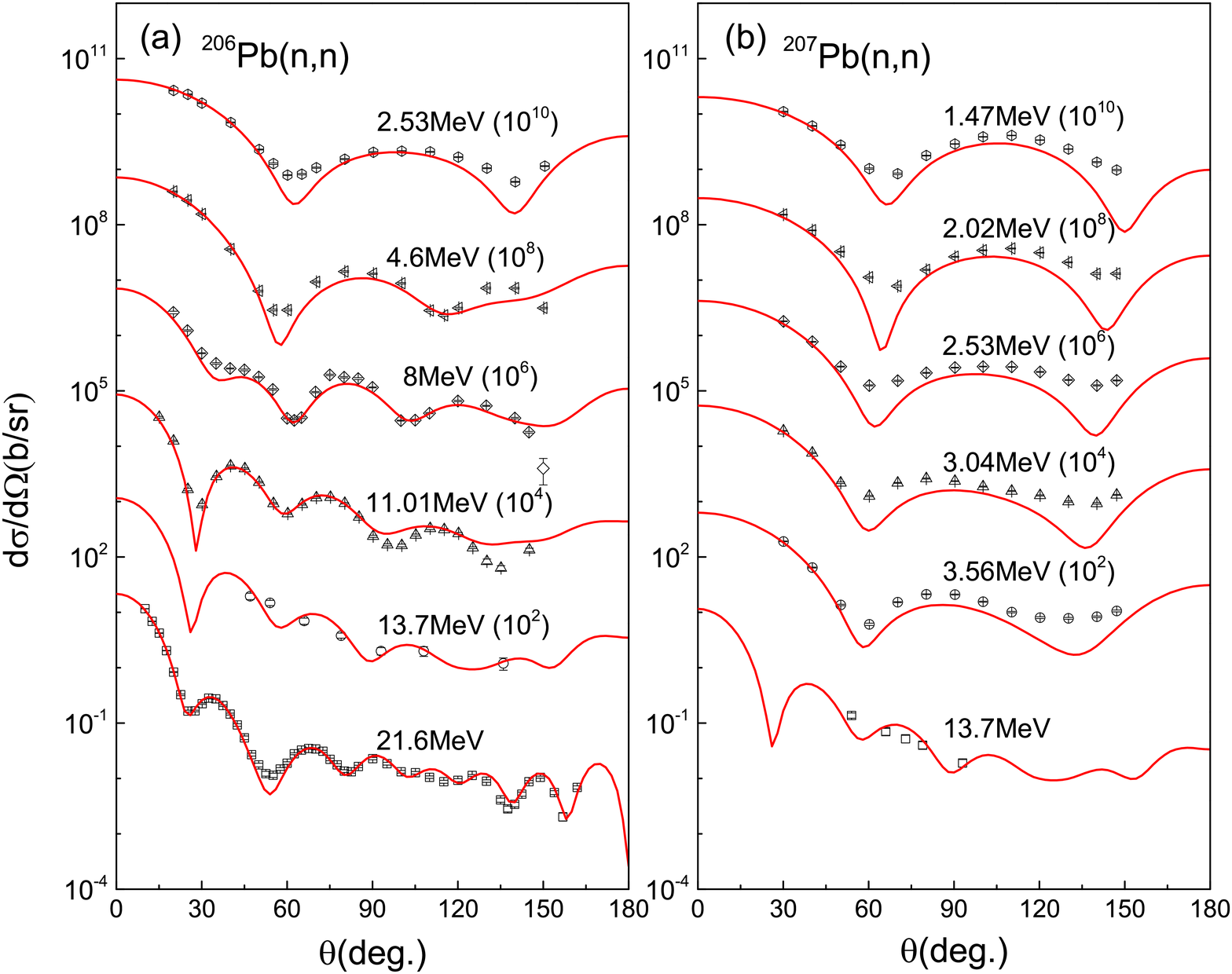}}
\subfigure[~$^{209}$Bi (n,n) and (p,p) angular distributions.]
{\includegraphics[width=0.48\textwidth]{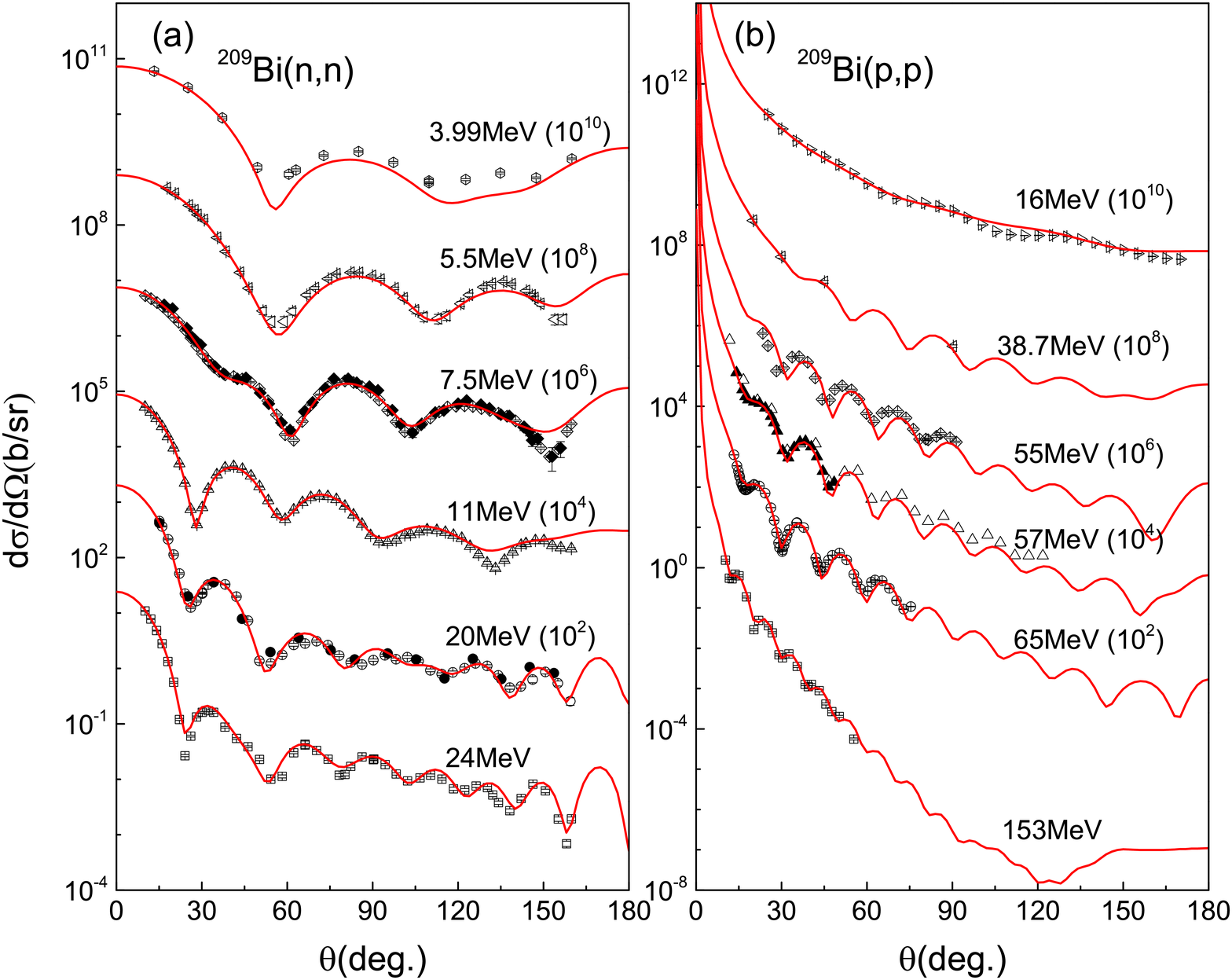}}
\vspace{-2mm}
\caption{Comparison of neutron and proton elastic scattering angular distributions with measurements at different incident nucleon energies.}
\label{fig6}
\vspace{-2mm}
\end{figure*}
\begin{figure*}[!tbh]
\vspace{-2mm}
\centering
\subfigure[~$^{208}$Pb elastic scattering analyzing powers.]
{\includegraphics[width=0.47\textwidth]{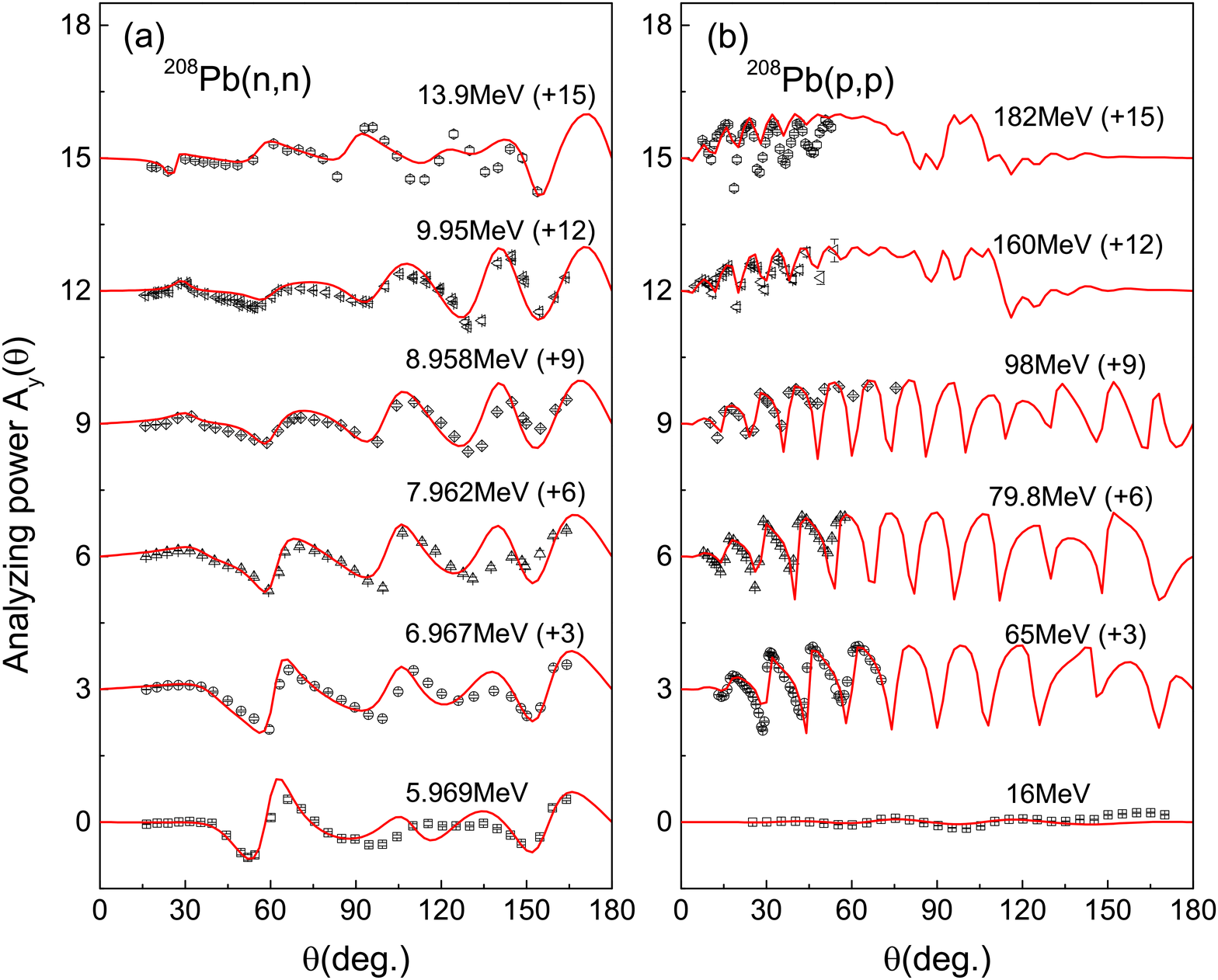}}
\subfigure[~$^{209}$Bi elastic scattering analyzing powers.]
{\includegraphics[width=0.47\textwidth]{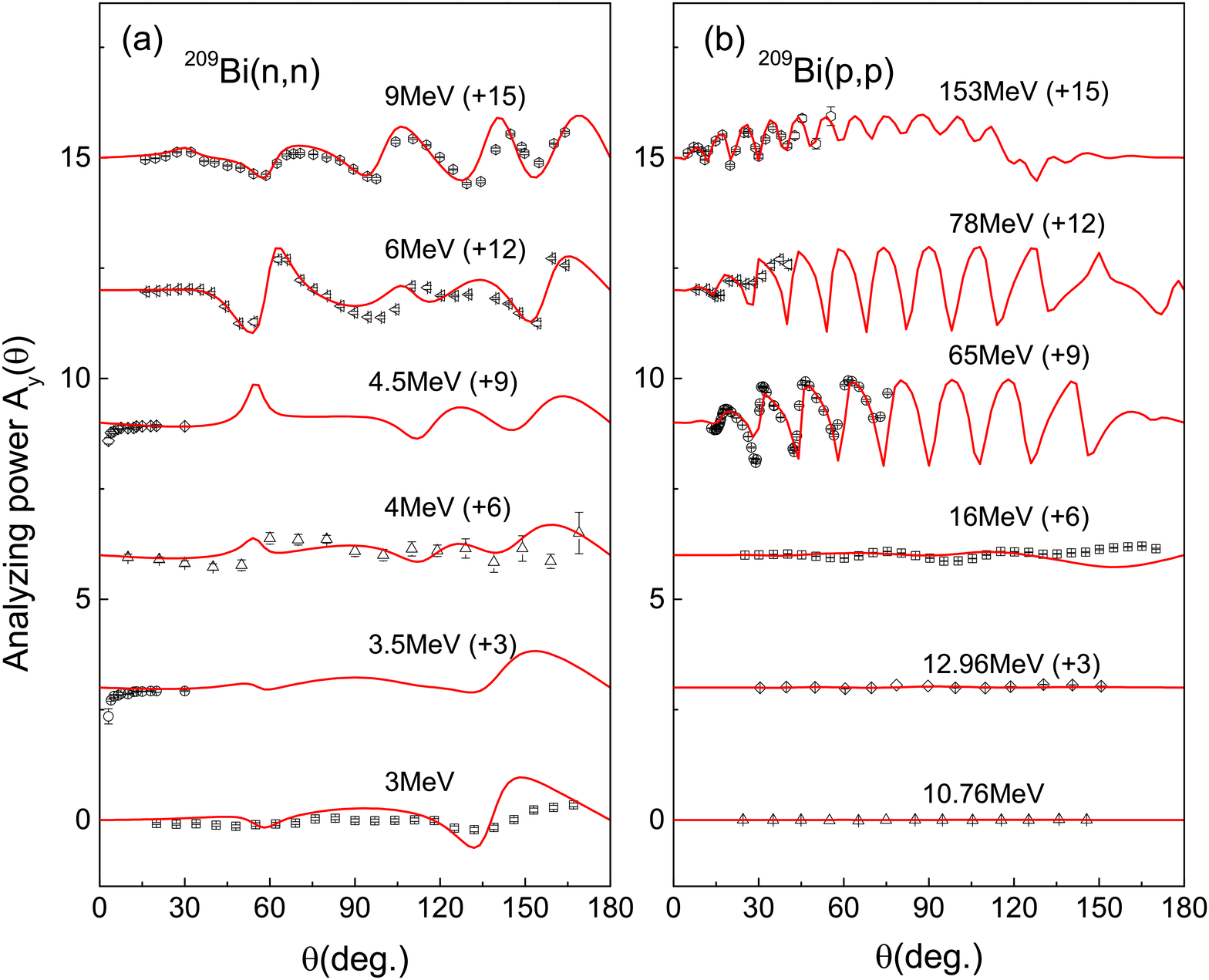}}
\vspace{-2mm}
\caption{Nucleon elastic scattering analyzing powers compared with the experimental data at different incident nucleon energies.}
\label{fig7}
\vspace{-3mm}
\end{figure*}
\begin{figure}[!tbh]
\centering
\subfigure[~$^{208}$Pb quasi-elastic (pn) scattering angular distributions.]
{\includegraphics[width=0.48\textwidth]{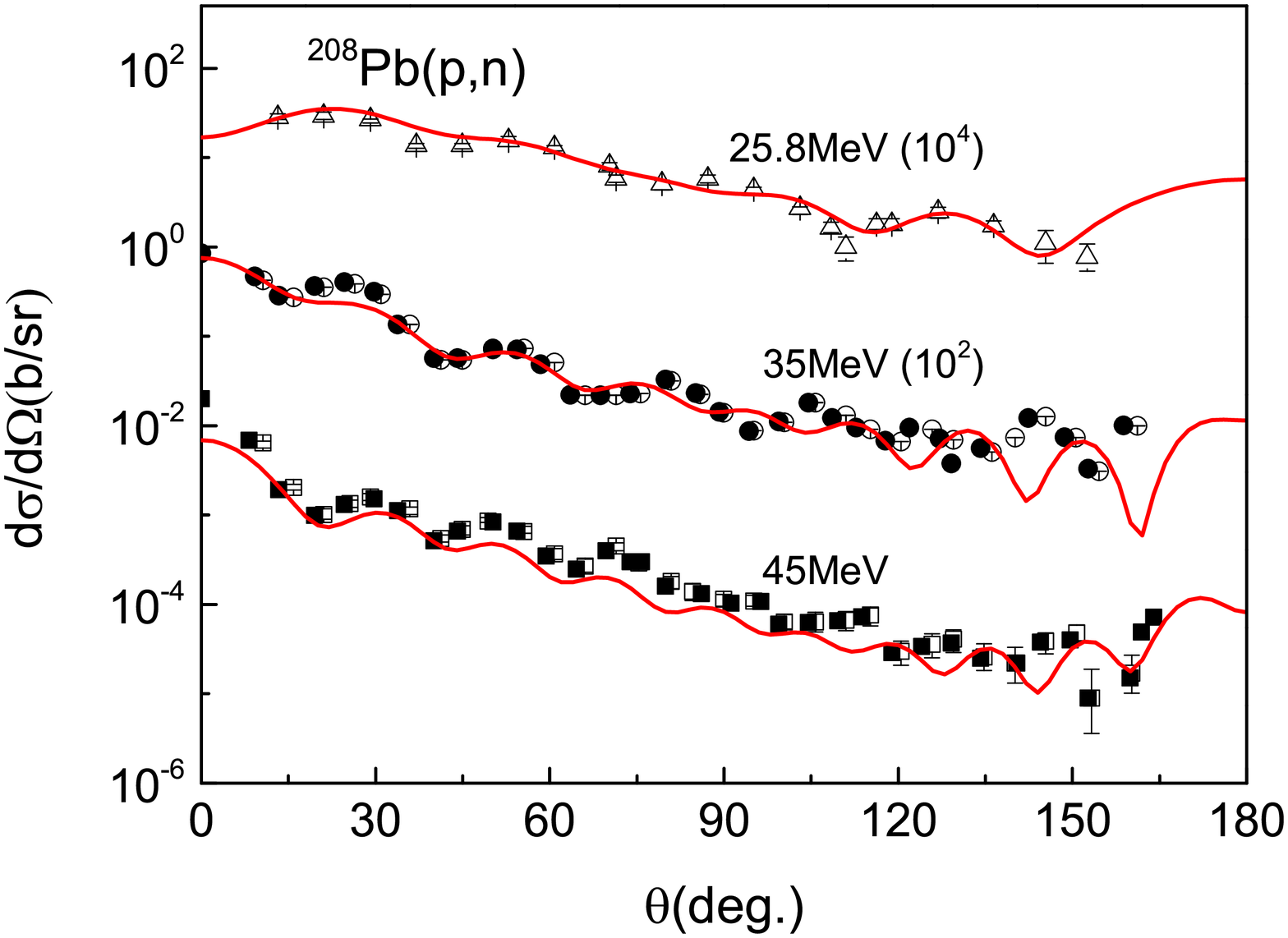}}
\subfigure[~$^{206}$Pb and $^{209}$Bi quasi-elastic (pn) scattering angular distributions.]
{\includegraphics[width=0.48\textwidth]{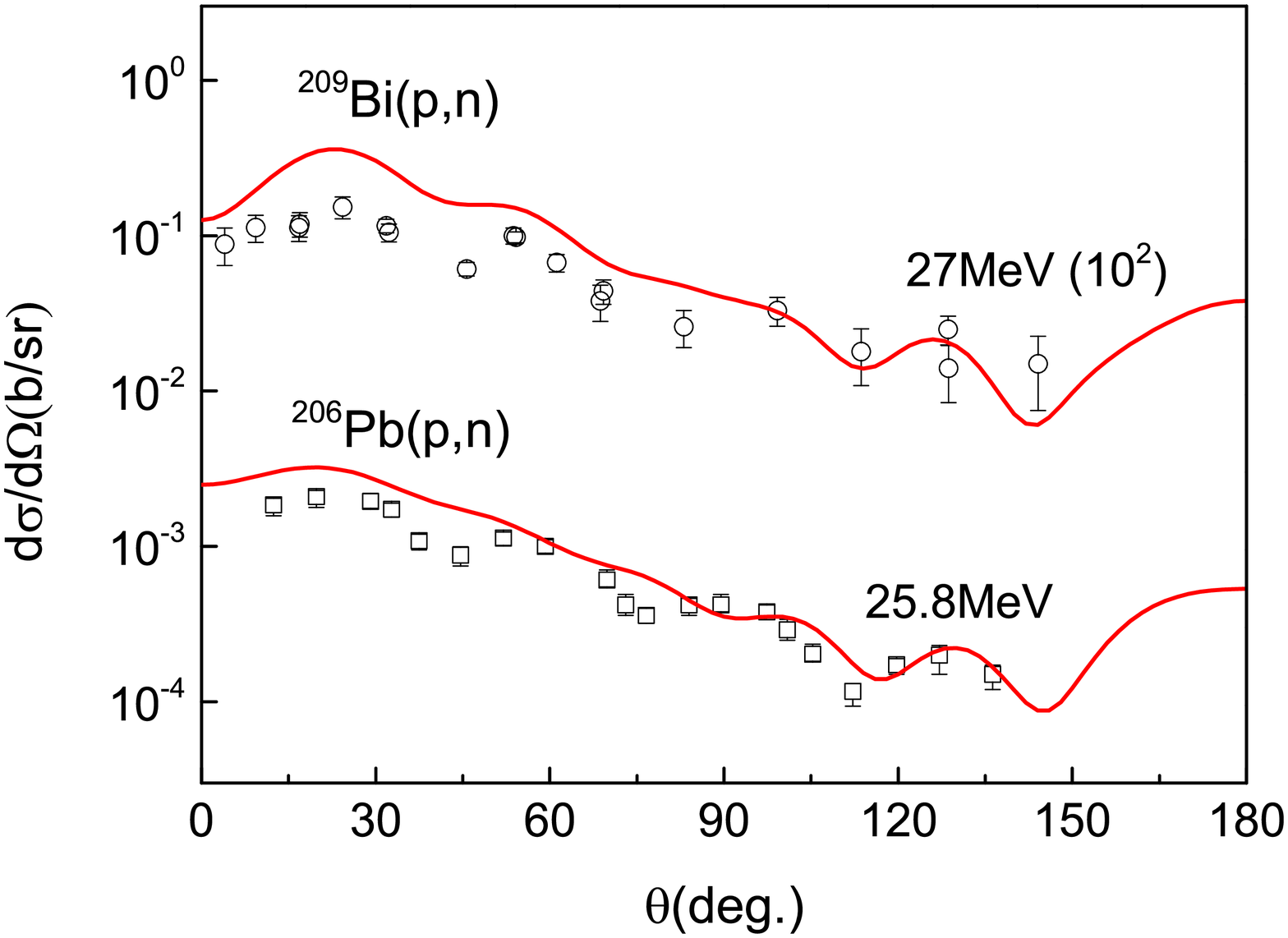}}
\vspace{-2mm}
\caption{Calculated angular distributions of the quasi-elastic (pn) scattering on $^{206}$Pb, $^{208}$Pb and $^{209}$Bi targets.}
\label{fig8}
\vspace{-4mm}
\end{figure}

Optical model code OPTMAN \cite{OPTMAN1,OPTMAN2,OPTMAN3} that includes the calculation of ($p$,$n$) quasi-elastic scattering~\cite{Quesada:2007} was used for cross-section calculations for positive energies.
The parameters of the dispersive optical model potential were searched for by minimizing the quantity $\chi ^2$ in the usual way \cite{SRM-ni:00}. All experimental data used in the fitting process are taken from the EXFOR database \cite{EXFOR} and is exactly the same database used to derive the DOMPs describing scattering on \PB~target and published in Refs.~\cite{Sun17,Zhao:2019}.

Additionally, the calculation of \PB~bound states that depends on the real potential~\cite{Zhao:2019} is also used in the DOMP optimization using the experimental data quoted in Ref.~\cite{johoma87}. Newly derived DOMP parameters are listed in Table \ref{tabI} and corresponding average particle (hole) energies $E_{\rm{P}}$(n) and $E_{\rm{P}}$(p) that define the imaginary potentials are listed in Table \ref{tabII}.

Figure \ref{fig1}(a) shows the obtained energy dependence of the real spin-orbit potential, of the imaginary (absorptive) potentials, and of the corresponding dispersive correction terms near the Fermi energy for the n+$^{208}$Pb reaction. A comparison of the energy dependence of the \ql Hartree-Fock\qr $V_{HF}$ potential is shown in Fig.~\ref{fig1}(b). \ql Previous\qr refers to the $V_{HF}$ potential from Ref.~\cite{Zhao:2019} which is compared to the local approximation of the non-local potential used in this work (see Eq.~\eqref{PB-approx} labelled as \ql Present\qrs).

The depth of new \ql Hartree-Fock\qr $V_{HF}$ potential given by Eq.~\eqref{PB-approx} is lower below the Fermi energy, falls more slowly up to 100 MeV and decreases faster above that energy as compared to the exponentially decreasing potential used in Ref.~\cite{Zhao:2019}. A shallower potential well given by the Perey-Buck non-local approximation \cite{PereyBuck62} proved to improve the description of the bound-states as well as scattering data as will be shown below.

Figure~\ref{fig2} zooms on the energy dependence of the imaginary (absorptive) potentials and corresponding dispersive-correction terms near the Fermi energy for the n+$^{208}$Pb system. The figure clearly shows that the imaginary potentials vanish from the energy ($2E_F-E_P$) up to the energy $E_P$ reflecting the shell gap. However, the dispersive correction remains non-zero in that region as discussed in Refs.~\cite{CPC,PRC-disp}.

\section{Results and discussion}
The calculation of neutron total cross section for the $^{208}$Pb target using our new DOMP is compared with the results of Koning-Delaroche \cite{Koning03} and our previously derived DOMP \cite{Zhao:2019} in Fig.~\ref{fig3} in the energy range from 6 up to 16~MeV. The potential from Ref.~\cite{Zhao:2019}  was worse than Koning-Delaroche description \cite{Koning03} in this region. Results from the current work shows a clear improvement over our previous work, the new DOMP results are in good agreement with data as well as with Koning-Delaroche potential calculations in this energy region.

The calculation of neutron total cross sections for $^{206}$Pb, $^{207}$Pb, $^{208}$Pb and $^{209}$Bi are compared in Fig.~\ref{fig4} with the results of Koning-Delaroche potential  \cite{Koning03} from 500 keV up to 200 MeV of incident neutron energy. The calculated total cross section using the new DOMP is in fair agreement with Koning-Delaroche results above 5 MeV, but reproduces better the experimental data below that energy for all targets.

The real part of our derived DOM potential is the shell model potential, and can be used to calculate the energies of the bound single-particle states of the magic nuclei $^{208}$Pb. This potential includes the sum of the Hartree-Fork term V$_{HF}$(E$_{nlj}$), the real spin-orbit term V$_{SO}$(E$_{nlj}$), and all dispersive correction terms $\Delta$V$_{v}$(E$_{nlj}$), $\Delta$V$_{s}$(E$_{nlj}$) and $\Delta$V$_{so}$(E$_{nlj}$) with the corresponding geometry-form-factors.

The experimental values of the neutron single-particle energies of the various single-particle and hole states for $^{208}$Pb were taken from  Ref.~\cite{johoma87}. The predicted single particle (hole) energies are compared with the experimental data in Fig.~\ref{fig5}. Results labelled \ql DOM(MR)\qr and \ql DOM(MR+35$\%$)\qr represent the Morillon and Romain calculations from Ref.~\cite{moro:06}; the label \ql $DOM_{previous}$\qr corresponds to calculations from our previous publication~\cite{Zhao:2019}, and the label \ql $DOM_{present}$\qr represents the current work. The description of the single particle bound states is significantly improved compared to Ref.~\cite{moro:06}, and slight improvement can be seen relative to our previous work. The order of both particle and hole states agrees with the experimental one; the particle energies agree well for the 5 single-particle levels near the Fermi energy; the agreement deteriorates for more un-bound states. A similar situation is observed for hole states - better agreement near the Fermi energy, worse for deeper hole states.

The neutron single-particle energies for last single-particle state and first single-hole state were calculated for the \PB~target; the absolute values of these two energies define the neutron separation energies S$_{n}$(A) and S$_{n}$(A+1). The calculated values of S$_{n}$(A) and  S$_{n}$(A+1) are 7.47~MeV and 3.85~MeV, respectively. These results are in excellent agreement with the corresponding experimental data 7.37~MeV and 3.94~MeV~\cite{Martin07,Chen15}. The root mean square(rms) radii for each orbit and single particle densities were also calculated and the agreement with results from Ref.~\cite{johoma87} is similar to what we already published for $^{208}$Pb \cite{Zhao:2019}.

The spectroscopic factor is given by the following expression,
\begin{eqnarray}
S_{nlj}&&=\int \bar{u}_{nlj}^{2}(r)[m/\bar{m}(r;E_{nlj})]dr\\
       &&=\int u_{nlj}^{2}(r)\frac{[m^{\ast}_{H}(r;E_{nlj})/m]}{[\bar{m}(r;E_{nlj})/m]}dr\\
       &&=\int u_{nlj}^{2}(r)\frac{[1-\frac{d}{dE}V_{HF}(r;E)\big|_{E=E_{nlj}}]}{[1-\frac{d}{dE}\Delta V(r;E)\big|_{E=E_{nlj}}]}dr.
\end{eqnarray}
where $\bar{u}_{nlj}$ is the eigenstate of the full microscopic mean field, and $u_{nlj}$ is the eigenstate of its local equivalent that we use.
Normalized eigenstates were used in spectroscopic factor calculations; details of the definition can be found in Refs~\cite{johoma87,Dieperink82}.
The spectroscopic factors of valence neutron particle and hole states  in $^{208}$Pb are compared in Table~\ref{tabIII} with previous calculations from Refs.~\cite{johoma87,Li81,Perazzo80,Bernard80,Hamamoto76,Ring73} (values were taken from Ref.~\cite{masa91rev}, except Johnson \etal values \cite{johoma87}). A reasonable agreement is observed.

\begin{table*}[bht]
\vspace{-2mm}
\caption{Spectroscopic factors of valence neutron particle (left half) and hole (right half) states in $^{208}$Pb.}
\vskip 0mm \renewcommand{\arraystretch}{1.25} \tabcolsep 5pt
\centerline{\begin{tabular*}{0.88\textwidth}{ccccccc|ccccccc}
\hline
\hline
 &3d$_{3/2}$&2g$_{7/2}$&4s$_{1/2}$&3d$_{5/2}$&1j$_{15/2}$&1i$_{11/2}$&2g$_{9/2}$ &3p$_{1/2}$&2f$_{5/2}$&3p$_{3/2}$&1i$_{13/2}$&2f$_{7/2}$&1h$_{9/2}$\\
\hline
this work            &0.90 &0.89&0.92&0.90&0.73&0.74&0.78&0.83&0.77&0.86&0.70&0.93&0.97\\
Ref~\cite{johoma87}  &0.90 &0.86&0.91&0.88&0.82&0.82&0.81&0.80&0.81&0.81&0.81&0.85&0.85\\
Ref~\cite{Li81}      &0.72 &0.80&0.61&0.77&0.55&0.73&0.67&0.79&0.73&0.71&0.62&0.53&0.51\\
Ref~\cite{Perazzo80} &0.79 &0.79&0.81&0.78&0.79&0.79&0.88&0.78&0.74&0.82&0.83&0.70&0.81\\
Ref~\cite{Bernard80} &0.80 &0.82&0.86&0.83&0.71&0.75&0.86&0.73&0.82&0.76&0.72&0.59&0.44\\
Ref~\cite{Hamamoto76}&0.72 &0.76&0.80&0.78&0.66&0.86&0.94&0.94&0.91&0.94&0.92&0.75&0.82\\
Ref~\cite{Ring73}    &0.88 &0.92&0.83&0.90&0.62&0.89&0.97&0.95&0.92&0.94&0.87&0.70&0.84\\
\hline
\end{tabular*}}
\label{tabIII}
\vspace{-3mm}
\end{table*}

Figure~\ref{fig6} shows calculated elastic scattering angular distributions of neutrons and protons incident on $^{206}$Pb, $^{207}$Pb, and $^{209}$Bi for different incident nucleon energies. Results for $^{208}$Pb are similar to those presented in Ref.~\cite{Zhao:2019} and are not shown in this paper.
The results for both neutron and proton elastic scattering describe the experimental data rather well over the entire energy and angular range. Slight underestimation of data below 5~MeV of neutron incident energy is probably associated with the missing compound-elastic contribution. 

Figure~\ref{fig7} shows elastic scattering analyzing powers of neutrons and protons incident on $^{208}$Pb and $^{209}$Bi for different incident nucleon energies. Results for $^{208}$Pb are similar to those presented in Ref.~\cite{Zhao:2019} in the regions where data are available; the agreement is reasonable but not perfect. Similar level of agreement is observed for nucleon scattering on $^{209}$Bi.

The newly fitted DCCOM potential has not been tested on quasielastic (p,n) scattering to the isobaric analog states (IAS) of the target nucleus. Such calculations represent the best
test of the isovector part of the optical potential. Figure \ref{fig8} shows the calculated quasi-elastic (p,n) angular distribution for scattering on $^{206}$Pb, $^{208}$Pb and $^{209}$Bi targets. Reasonable agreement with data is achieved showing the Lane consistency of the derived DOMP, \textit{i.e.}, the same potential describes both neutron and proton scattering indistinctly, including the quasi-elastic (p,n) scattering which is defined by the isovector potential. However, additional work is needed to clarify a potential improvement of the quasi-elastic data description by introducing a shift of the isovector and isoscalar geometries as recently proposed by Danielewicz \textit{et al.} \cite{Dani:17}. In fact, our calculations underestimate the oscillations in data as observed in Ref.~\cite{Dani:17}.

\section{Conclusions}
In order to improve the description of the nucleon scattering data and the bound-state energies using a dispersive potential, this work considered the non-locality of the real potential as suggested by Perey and Buck~\cite{PereyBuck62} and extensively used in papers by CEA Bruy\`{e}res-le-Ch\^{a}tel and Washington University (St Louis) groups, and the impact of the large shell gap in magic nuclei on the definition of the imaginary potential~\cite{masa91rev,mocaqu:02}. The improved physical model allowed to derive a Lane-consistent dispersive optical model potential that accurately describes scattering data for nucleon induced reactions on the double-magic target $^{208}$Pb. The real part of the same DOMP, which corresponds to the shell potential, gives a good description of the bound-state data. Newly derived potential is also shown to give a good description of  nucleon scattering on near-magic lead and Bi isotopes, which is very important for applications.

\section*{ACKNOWLEDGEMENT}
This work is partly supported by Science Challenge Project, No. TZ2018001 and TZ2018005.

\end{document}